\documentclass[12pt]{iopart}
\usepackage{iopams}
\usepackage{graphicx}
\usepackage{bm}
\usepackage{color}
\usepackage{dsfont}

\begin{document}

\def\bra#1{\langle{#1}|}
\def\ket#1{|{#1}\rangle}
\def\sinc{\mathop{\text{sinc}}\nolimits}
\def\cV{\mathcal{V}}
\def\cH{\mathcal{H}}
\def\cT{\mathcal{T}}
\renewcommand{\Re}{\mathop{\text{Re}}\nolimits}

\def\e{\mathrm{e}}
\def\ii{\mathrm{i}}
\def\d{\mathrm{d}}

\definecolor{dgreen}{rgb}{0,0.5,0}
\newcommand{\green}{\color{dgreen}}
\newcommand{\RED}[1]{{\color{red}#1}}
\newcommand{\BLUE}[1]{{\color{blue}#1}}
\newcommand{\GREEN}[1]{{\color{dgreen}#1}}
\newcommand{\REV}[1]{{\color{red}[[#1]]}}
\newcommand{\SP}[1]{{\color{blue} [[#1. Saverio]]}}
\newcommand{\rev}[1]{{\color{red} #1}}

\def\HN#1{{\color{magenta}#1}}
\def\DEL#1{{\color{yellow}#1}}

\title[Discrete Abelian Gauge Theories for Quantum Simulations of QED]{Discrete Abelian Gauge Theories for Quantum Simulations of QED}

\author{Simone Notarnicola$^{1}$, Elisa Ercolessi$^{2,3}$, Paolo Facchi$^{4,5}$, Giuseppe Marmo$^{6,7}$, Saverio Pascazio$^{4,5}$, Francesco V. Pepe$^{4,5,8}$}

\address{$^1$SISSA, International School for Advanced Studies, via Bonomea 265, I-34136 Trieste, Italy}
\address{$^2$Dipartimento di Fisica e Astronomia dell'Universit\`a di Bologna, Via Irnerio 46, I-40127 Bologna, Italy}
\address{$^3$INFN, Sezione di Bologna, Via Irnerio 46, I-40127 Bologna, Italy}
\address{$^4$Dipartimento di Fisica and MECENAS, Universit\`a di Bari, I-70126 Bari, Italy}
\address{$^5$INFN, Sezione di Bari, I-70126 Bari, Italy}
\address{$^6$Dipartimento di Fisica and MECENAS, Universit\`a di Napoli ``Federico II'', I-80126 Napoli, Italy} 
\address{$^7$INFN, Sezione di Napoli, I-80126 Napoli, Italy}
\address{$^8$Museo Storico della Fisica e Centro Studi e Ricerche ``Enrico Fermi'', I-00184 Roma, Italy}

\begin{abstract}
We study a lattice gauge theory in Wilson's Hamiltonian formalism. In view of the realization of a quantum simulator for QED in one dimension, we introduce an Abelian model with a discrete gauge symmetry $\mathbb{Z}_n$, approximating the $U(1)$ theory for large $n$. We analyze the role of the finiteness of the gauge fields and the properties of physical states, that satisfy a generalized Gauss's law. We finally discuss a possible implementation strategy, that involves an effective dynamics in physical space.
\end{abstract}


\vspace{2pc}


\section{Introduction}

Recent development in low-temperature physics and atomic control
techniques is providing the basic tools for setting up quantum
simulators \cite{qsim1,qsim2,qsim3,qsim4}. The experimental
feasibility of a quantum simulator will open the way to a more
comprehensive understanding of complex systems and fundamental
physics. An appealing application is the simulation of lattice
gauge theories
\cite{rothe1992lattice,montvay1997quantum,wilsonlgt,kogut1975hamiltonian,susskind1977lattice,kogut1979introduction}.
The discretization of high-energy physics theories on lattices was
initially motivated by the possibility to simulate them by
classical computation. However, the complex nature of gauge
theories represents a severe obstruction, which can be overcome
through a \textit{quantum} simulation. Atoms on a lattice
\cite{simul1,simul2,simul3,simul4,simul5,simul6,simul7,simul8,simul9,simul10,simul11}
provide a natural toolbox to perform this task.

In this article we construct a lattice model which simulates an
Abelian gauge theory. The model will be characterized by the
interaction between an Abelian gauge (electromagnetic) field and a
fermionic matter field, as in quantum electrodynamics (QED). We shall
restrict our analysis to the one-dimensional case. As in the
consolidated quantum link model (QLM)
\cite{qlm1,qlm2,qlm3,qlm4wiese}, our system will be an
approximation to QED in which the electric field can take a
\textit{finite} number of values. We will eventually discuss the
possibility to implement the model on a cold atomic simulator.
This task involves first the identification of the degrees of
freedom of the simulator with those of the model, and then the
correct implementation of the dynamics.

\section{Lattice QED}
 
The dynamics of the continuum-space QED
in $1+1$ dimensions (Schwinger model) has its lattice counterpart
in the Hamiltonian \cite{montvay1997quantum,qlm4wiese,melnikov}
\begin{equation}\label{HamQED}
H = -t\sum_x \!\left(\psi^{\dagger}_x
U_{x,x+1}\psi_{x+1}+\mathrm{H.c.} \right)\! + m\sum_x(-1)^x\psi_x^{\dagger}\psi_x+\frac{g^2}{2}\sum_x
E_{x,x+1}^2,
\end{equation}
with $x$ labelling the sites of a one-dimensional lattice. 
Fermionic matter is represented by the one-component spinor field operators $\psi_{x}$, defined on each site,
which obey the canonical anticommutation relations
$\{\psi_{x},\psi_{x'}^{\dagger}\}=\delta_{x,x'}$, 
$\{\psi_{x},\psi_{x'}\}=0$.
The parameter $m$ is the fermion mass, while the staggered fermions with parity
factor $(-1)^x$ are introduced
in order to avoid the fermion-doubling problem in the 
discretization of the theory~\cite{kogut1975hamiltonian,susskind1977lattice,rothe1992lattice,montvay1997quantum}: the positive and negative-energy
components of the Dirac spinor are encoded respectively in the
even and odd lattice sites. In a simplified model, spinless
particles are considered, with the spinor $\psi_x$ reducing to a
single-component field. The gauge fields are instead defined on
the links $(x,x+1)$ of the lattice. In the canonical gauge, which
is the most convenient choice to develop a lattice gauge theory,
the electric field $E$ and the vector potential $A$ are
conjugated variables with canonical Commutation Relations (CR)
 $[E_{x,x+1},A_{x',x'+1}]=\ii\delta_{x,x'}$~\cite{fradkin}. Since the unitary operators $U$ (comparators
\cite{peskin1995introduction}) in the hopping terms of~(\ref{HamQED}) 
are locally related to the vector potential by
exponentiation, $U_{x,x+1}=\e^{-\ii A_{x,x+1}}$, they satisfy
\begin{equation}\label{gaugecomm}
[E_{x,x+1},U_{x',x'+1}]=\delta_{x,x'} U_{x,x+1}.
\end{equation}
The presence of the free electric field energy, with coupling
constant $g^2$, thus yields a nontrivial dynamics for the
comparators. Notice the absence of magnetic contributions to the
Hamiltonian, a consequence of the one-dimensional nature of the
system.

The terms $\psi^{\dagger}_x U_{x,x+1} \psi_{x+1}$ 
in~(\ref{HamQED}) describe site hopping of fermions, related to a
shift in the electric field. These contributions come from the
discretization and integration on the lattice cells of the
minimal-coupling terms $-\ii\psi(x)^{\dagger}\gamma_0\gamma^j D_j
\psi(x)$,  
with $\gamma^{\mu}$ the Dirac matrices, and $D_j=\partial_j +i
A_j(x)$  the covariant derivatives~\cite{kogut1975hamiltonian,peskin1995introduction}. The minimal
coupling ensures the symmetry of the Hamiltonian under local
$U(1)$ transformations. Given a real function 
on the lattice, $\alpha_x$, local phase transformations
$\psi_x\rightarrow\psi_x \e^{\ii\alpha_x}$ and $U_{x,x+1}\rightarrow
\e^{\ii\alpha_x}U_{x,x+1} \e^{-\ii\alpha_{x+1}}$ of the field operators
leave the Hamiltonian (\ref{HamQED}) invariant. Due to the
(anti)commutation properties of the fields, the phase
transformation on any operator $F$ can be implemented through the
application of
\begin{eqnarray}
\label{Gx}
F &\to& \prod_x \e^{-\ii\alpha_x G_x} F \prod_y \e^{\ii\alpha_y G_y}, \\
 G_x &=& \psi_x^{\dagger}\psi_x + \frac{1}{2}[(-1)^x-1] -(E_{x,x+1}-E_{x-1,x}). \nonumber 
\end{eqnarray}
The gauge transformation acts trivially on the local-$U(1)$
invariant Hamiltonian (\ref{HamQED}). For a generic operator $F$,
the invariance property is equivalent to the CR $[F,G_x]=0$ for all
sites. The choice of the canonical gauge has the disadvantage that
Gauss's law cannot be implemented at the operator level
\cite{fradkin,qlm4wiese}. Thus, not all the states in the total
Hilbert space $\mathcal{H}$ of the system are physically
acceptable. The invariance condition of a state $|\phi\rangle$
under any gauge transformation $\prod_x \e^{-\ii\alpha_x G_x}
|\phi\rangle = |\phi\rangle$ selects the Hilbert subspace
\begin{equation}
\mathcal{H}_G = \left\{ |\phi\rangle \in \mathcal{H}, G_x
|\phi\rangle = 0 \mbox{ for all sites } x  \right\},
\end{equation}
where, as can be deduced from (\ref{Gx}), the charge
$\psi_x^{\dagger}\psi_x$ and the divergence of the electric field
$E_{x,x+1}-E_{x-1,x}$ are correctly related. The  
term multiple of the identity
appearing in $G_x$ ensures that the vacuum state with vanishing
electric field and all the negative-mass sites occupied (the
\textit{Dirac sea}) is in the gauge-invariant subspace
$\mathcal{H}_G$.

\section{Finite link spaces: a $\mathbb{Z}_n$ model.}

In Wilson's
original formulation of lattice gauge theories \cite{wilsonlgt},
the gauge operators on links act on infinite-dimensional Hilbert
spaces, and both the electric field and the vector potential have
continuum and unbounded spectra. A problem arises in quantum
simulators, when one has to match the (infinite dimensional) link  
with an experimentally feasible and
controllable system, with a \emph{finite} number of levels. 
Two possible approaches 
are possible. The first one preserves for all
dimensions the structure of the Hamiltonian (\ref{HamQED}),
including the coupling of the matter fields with a \emph{unitary} gauge
operator. The second approach, which has been followed in the
formulation of the QLM \cite{qlm1}, consists in
preserving the CR between field operators,
(\ref{gaugecomm}) in particular. Since in quantum mechanics one is
used to think in terms of commutators, the latter approach has so far appeared
more natural. Unfortunately, this procedure focuses on the invariance with respect to the U(1) group of local transformations, at the expenses of the structure of the hopping term, that no longer involves a unitary comparator (minimal coupling prescription). We shall rather insist on the unitarity of the comparator thereby obtaining a \textit{bona fide} lattice \emph{gauge} theory
for any dimension of the link Hilbert space. Let us first
observe that a gauge transformation (\ref{Gx}) acts on the
comparator $U_{x,x+1}=\e^{-\ii A_{x,x+1}}$ as
\begin{equation}\label{trasfcomp}
U_{x,x+1} \rightarrow \e^{\ii(\alpha_x -\alpha_{x+1})E_{x,x+1}}
U_{x,x+1} \e^{-\ii(\alpha_x-\alpha_{x+1}) E_{x,x+1}}
= \e^{\ii(\alpha_x-\alpha_{x+1})} U_{x,x+1}.
\end{equation}
This result is indeed a special case of a general property of
operators in the Weyl group generated by the conjugated operators
$A$ and $E$ on a link (indices will be omitted for clarity)
\cite{weyl1950theory,schwinger2001quantum}. Indeed, the electric
field and the vector potential are the generators of the
two-parameter projective unitary (Weyl) group $\left\{ \e^{\ii(\xi E-\eta A)}\right\}_{\xi,\eta \in \mathbb{R}}$.
Using the canonical CR $[E,A]=\ii$ for the
generators, the following relation holds for any $\xi$ and $\eta$
\begin{equation}\label{WeylComm}
\e^{\ii\xi E} \e^{-\ii\eta A} \e^{-\ii\xi E} = \e^{\ii\eta
\xi} \e^{-\ii\eta A},
\end{equation}
which particularizes to (\ref{trasfcomp}) for $\eta=1$ and $\xi=\alpha_x-\alpha_{x+1}$.

In a finite-dimensional Hilbert space, the role of generators
loses its meaning. Nonetheless, we can define a set of unitary
operators, the (discrete) Schwinger-Weyl group, that satisfy the relation~(\ref{WeylComm}). 
Notice that this entails a change of paradigm. We are abandoning an approach in terms of the algebra of generators in favor of an alternative approach in terms of its group, enabling us to explore the global features of the topology. This concept is elaborated in the Appendix. Observe
that the operator $\e^{-\ii \eta A}$ acts as a translation
of the electric field, since $\e^{\ii\eta A}E \e^{-\ii\eta A} = E+\eta$.
A similar role is played by the elements of
the Weyl group with $\xi=0$ with respect to $A$.

Let us now consider an $n$-dimensional Hilbert space and choose an
orthonormal basis $\{ |v_k \rangle \}_{1\leq k\leq n}$, 
which will be called the \textit{electric field basis}, and define a
unitary operator $U$ that performs a cyclic permutation of the
basis states:
\begin{equation}
\label{cyclic}
U |v_k\rangle = |v_{k+1}\rangle \quad \mbox{for } k<n, \quad
U|v_{n}\rangle = |v_{1}\rangle .
\end{equation}
We will call the orthonormal eigenbasis of $U$ the \textit{vector
potential basis}. The operator $V$ conjugated to $U$ is diagonal in the electric field basis, with $V\ket{v_k} = \e^{-\ii 2 \pi k/n} \ket{v_k}$, and it cyclically permutes the elements of the vector potential
basis. Since $U^n=V^n=\mathbb{I}$, the sets $\{U^k\}_{1\leq k\leq
n}$ and $\{V^k\}_{1\leq k\leq n}$ are unitary representations of the group
$\mathbb{Z}_n$ of integers modulo $n$. The set of all the products between
$U$ and $V$ and their integer power constitutes the Schwinger-Weyl
group \cite{weyl1950theory,schwinger2001quantum,varadarajan}. The CR 
between the elements of this group yield the relation
\begin{equation}\label{commSW}
V^{-k} U^{\ell} V^k = \e^{\ii\frac{2\pi}{n}k\ell}U^{\ell} \quad
\mbox{with } k,\ell\in \mathbb{Z}.
\end{equation}
The multiplication law (\ref{commSW}) satisfied by the $\mathbb{Z}_n$ operators
is the discrete form of (\ref{WeylComm}), valid for $U(1)$
operators.

\begin{figure}
\centering
\includegraphics[width=0.75\textwidth]{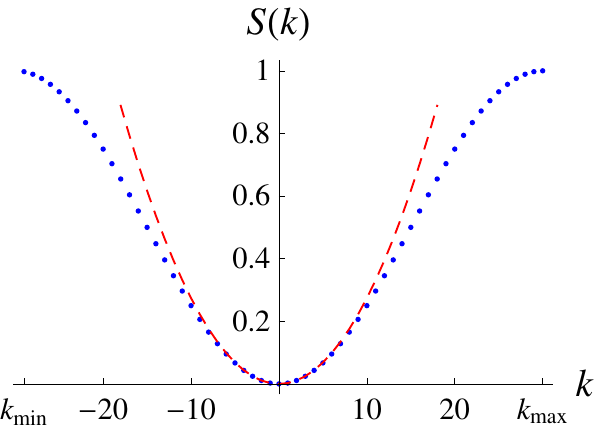}\\
\caption{Spectrum $S(k_{x,x+1})$ of the operator $f(V_{x,x+1})$
defined in (\ref{fV}). The dashed (red) line represents the
quadratic approximation of the spectrum close to its minimum at
$k=0$.}
\label{imm:cfr}
\end{figure}

\begin{figure}
\centering
\includegraphics[width=0.75\textwidth]{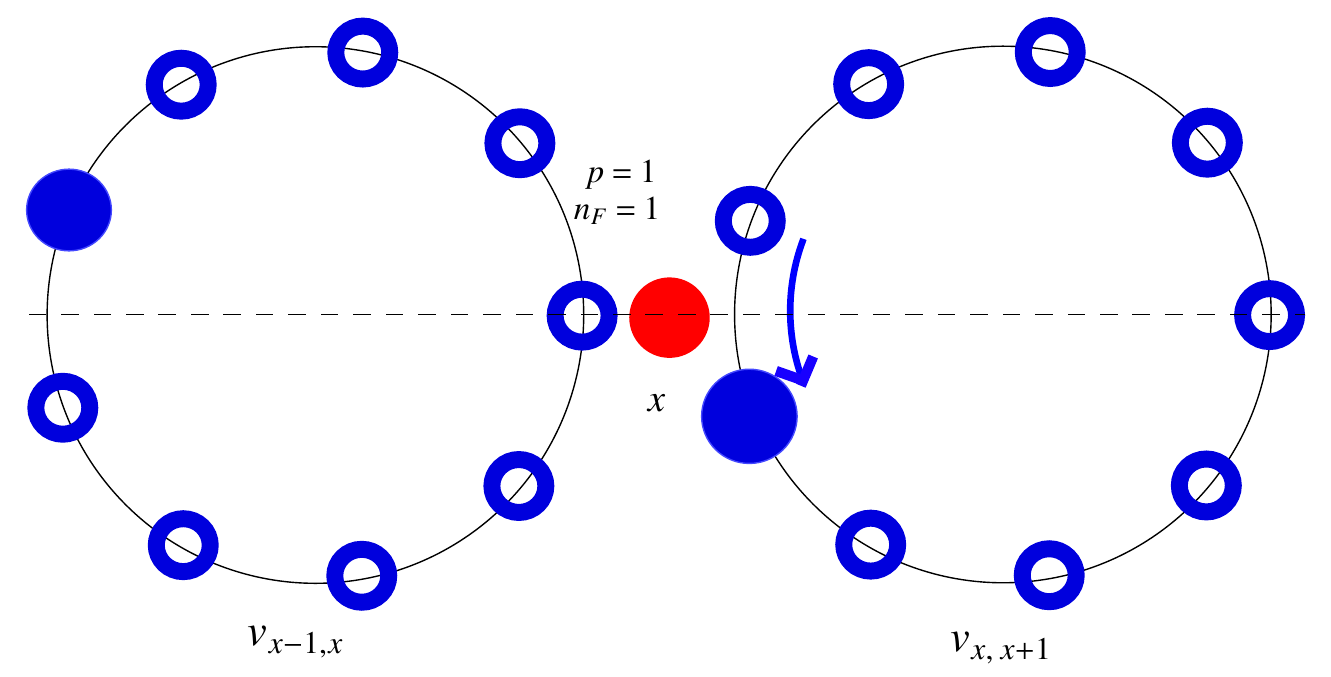}
\caption{
The state of two links (large circles) neighboring an
even site [central full dot, $p=(-1)^x=1$].
In this example, the site is occupied.
A full dot on the circle represents the occupied eigenstate of
$V$, while the empty dots are the other eigenstates.
Observe that unitarity 
entails probability conservation: the full dot never leaves the circle under (\ref{cyclic}).}
\label{fig:1imm}
\end{figure}

Once the correspondences $U_{x,x+1}\leftrightarrow
\e^{-\ii A_{x,x+1}}$ and $V_{x,x+1}\leftrightarrow \e^{-\ii E_{x,x+1}}$ has
been set up for all links, we can construct an Abelian theory
which represents an approximation to the lattice QED Hamiltonian~(\ref{HamQED}) with a local $\mathbb{Z}_n$ invariance \cite{zohar2015,zoharreview}. One
of the earliest examples of a pure gauge model with a $\mathbb{Z}_n$
invariance, used to approximate a $U(1)$ theory in the
$n\to\infty$ limit, was given in \cite{elitzur}. The dynamics of
the new model is determined by the following variant 
of the Hamiltonian~(\ref{HamQED})
\begin{equation}\label{Ham}
H_n = -t\sum_x (\psi^{\dagger}_x U_{x,x+1}\psi_{x+1} +
\mathrm{H.c.}) + m \sum_x (-1)^x
\psi_x^{\dagger}\psi_x + \frac{g_n^2}{2}\sum_x f(V_{x,x+1}),
\end{equation}
where 
$f(V)$ is
a suitable Hermitian operator, diagonal in the electric field basis,
which represents the discretized free-field electromagnetic
Hamiltonian. Unlike in the QLM, the
ladder operators $U$ permute the electric field basis states on a
circle, and transitions between neighboring states all occur with
the same amplitude. These features are a consequence of the request that the
minimal-coupling structure in (\ref{HamQED}) is preserved in its
finite-dimensional link counterpart. The function $f(V)$ is so far arbitrary. 
We shall introduce a function $f(V)$ that, like $E^2$ in (\ref{HamQED}), has a single minimum. A simple
choice is
\begin{equation}\label{fV}
f(V_{x,x+1}) = \frac{1}{4} \bigl( V_{x,x+1}-\mathbb{I} \bigr) \bigl(
V_{x,x+1}^{\dagger} - \mathbb{I} \bigr).
\end{equation}
Since the eigenvalues of $V$ are 
$v_k = \e^{-\ii  2\pi k/n}$,
the operator~(\ref{fV}) has eigenvalues
$S(k_{x,x+1}) = ( \sin \frac{\pi k_{x,x+1}}{n} )^2$.
For the low-energy states around the minimum at $k=0$, the
spectrum of $f(V)$ is quadratic like the energy $E^2$ associated
to the electric field in the original model. See Fig.~\ref{imm:cfr}.
We remark that the choice of a function for the electric field energy
does not affect the discretization of the gauge
theory, as long as it is characterized by a single minimum. The energy term (\ref{fV}) is characterized by a nondegenerate ground state. Another interesting possibility is obtained by the replacement $V\to \e^{-\ii\pi/n}V$, yielding two degenerate minima at $k=-1$ and $k=0$, which could be of interest for studying the effects of terms that break the chiral symmetry of the gauge field \cite{coleman1,coleman2,simul2}.
As expected, in the link discretization process the $U(1)$ gauge
symmetry of the original theory becomes a $\mathbb{Z}_n$ symmetry.
Indeed, due to condition (\ref{commSW}), the Hamiltonian $H_n$ is
invariant under the transformation
\begin{equation}\label{trasfHn}
H_n \to \prod_x (T_x^{\dagger})^{\nu_x} H_n \prod_y (T_y)^{\nu_y},
\end{equation}
where $\nu_x$ is an arbitrary integer-valued function on the lattice, and
\begin{equation}\label{Tx}
T_x = \e^{\frac{2\pi \ii}{n} \!\left(\psi_x^\dagger\psi_x +
\frac{(-1)^x-1}{2}\right)} V_{x,x+1} V^{\dagger}_{x-1,x},
\end{equation}
with $(T_x)^n=\mathbb{I}$, represents the discretized analogue in
the Schwinger-Weyl theory of $\e^{\ii G_x}$ [see
Eq.~(\ref{Gx})], obtained by the correspondence 
$\e^{-\ii E}\to V$. Due to the arbitrariness of $\nu_x$, the
gauge-invariance condition is equivalent to
\begin{equation}\label{gaugedyn}
[H_n,T_x]=0, \qquad \forall x.
\end{equation}
In the $\mathbb{Z}_n$ theory, the Hilbert subspace of physical states is
determined by a generalized Gauss's law:
\begin{equation}\label{physpace}
\mathcal{H}_T = \left\{ |\phi\rangle \in \mathcal{H} ,\, T_x
|\phi\rangle = |\phi\rangle \mbox{ for all sites } x 
\right\}.
\end{equation}
In the \emph{reference basis} $\left\{ |n_x,v_{x,x+1} \rangle \right\}$,
determined by the eigenvalues of $V_{x,x+1}$ and of
$\psi_x^{\dagger} \psi_x$, namely the occupation numbers
$n_x\in\{0,1\}$, the gauge invariance request translates
into a condition on the eigenvalues. Since each $T_x$ acts
nontrivially on site $x$ and the adjacent links, the physical
subspace (\ref{physpace}) is spanned by the reference basis
states satisfying
\begin{equation}
\e^{\frac{2\pi \ii}{n} \!\left(n_x + \frac{(-1)^x-1}{2}\right)}
v_{x,x+1} v^{*}_{x-1,x} = 1, \qquad \forall x.
\end{equation}
If an even site with $(-1)^x=1$ is empty, the
eigenvalues of $V$ in neighboring links must be equal, while
if it is occupied, they will be related by $v_{x,x+1}=\e^{-2\pi \ii/n}
v_{x-1,x}$. On the other hand, in odd sites with $(-1)^x=-1$, the
eigenvalues of $V$ are equal if the site is occupied, while
they are related by $v_{x,x+1} = \e^{2\pi \ii/n} v_{x-1,x}$ otherwise. This
is in agreement with Dirac's picture in which the absence of a particle
in a negative-energy site is equivalent to the presence of an
antiparticle. We can represent these
situations by visualizing the eigenstates of each $V$ as $n$
points placed at an angular distance of $2\pi/n$ on a circle. 
Figure \ref{fig:1imm} displays the state of an occupied even site. The action of the operator $U_{x,x+1}$
($U_{x,x+1}^{\dagger})$ induces a counterclockwise (clockwise) jump of
the eigenstate of $V$ to a neighboring state on the circle. The
correlation between the jump in the link eigenstate and the charge
displacement in the terms $\psi_x^\dagger U_{x,x+1}\psi_{x+1}$
(and conjugates) keeps the system in the physical subspace
$\mathcal{H}_T$.

Although the approximation of Wilson's model with a $\mathbb{Z}_n$ lattice gauge theory improves with increasing $n$, interesting phenomenology emerges also at small link dimension: the $\mathbb{Z}_3$ theory is already a good test bed for the analysis of electric flux string breaking \cite{simul2,montangero2015}. Moreover, $\mathbb{Z}_n$ theories are relevant \textit{per se}, being related to the problem of confinement in QCD \cite{zoharreview}.

Let us finally remark that an extension to higher dimensional lattices would involve a plaquette term in
the Hamiltonian with a product of four $U$ matrices
\cite{qlm4wiese}. The $\mathbb{Z}_n$ theory can be generalized by a proper
modification of  Gauss's law, which takes into account all the
terms in the divergence of the electric field.

\section{Implementation strategy}

The implementation of the
Hamiltonian (\ref{Ham}) is a difficult task due to the presence of
the correlated hopping terms $\psi_x^{\dagger} U_{x,x+1}
\psi_{x+1}$, describing elementary processes in which the hopping $x \to x+1$
of a fermion to a nearest-neighboring site is always associated with an electric-field change on the link $(x,x+1)$ between the two sites.  
Recent results indicate that it is
possible to engineer a simpler Hamiltonian, and then obtain the
gauge theory by imposing the gauge-invariance constraint(s) by
assigning an energy penalty to the non gauge-invariant subspace
\cite{simul2,zohar2001} or by implementing Zeno constraints
\cite{simul11,zeno1,zeno2,zeno3,zeno4}. We shall follow the first
path, and induce the correlated hopping in (\ref{Ham}) through a
second-order effective gauge-invariant Hamiltonian.

In view of implementing a Hermitian term in the Hamiltonian, it is
more convenient to express the local gauge-invariance condition
$T_x |\phi\rangle= |\phi\rangle$ as
\begin{equation}
\Gamma_x |\phi\rangle = 0, \qquad 
\Gamma_x = (T_x-1)(T_x^{\dagger}-1), 
\end{equation}
so that the physical space is the kernel of
the positive operator
\begin{equation}
\Gamma = \sum_x \Gamma_x.
\end{equation}
The dynamical conservation of the gauge condition for all sites can
be enforced by adding a large term proportional to $\Gamma$ 
to the Hamiltonian. Notice that the constraint is diagonal in the
reference basis $\left\{ |n_x,v_{x,x+1} \rangle \right\}$.

Let us consider a Hamiltonian
that involves \textit{uncorrelated} hopping of fermions between
nearest-neighbor sites and transitions, with equal amplitudes,
between neighbor link states on a circle:
\begin{equation}\label{H0}
H_{(0)} = - \tilde{t} \sum_{x} \!\left(
\psi_{x}^{\dagger}\psi_{x+1} + \psi_{x+1}^{\dagger}\psi_{x}
\right)\! - \tilde{w} \sum_{x} \!\left( U_{x,x+1}
+ U_{x,x+1}^{\dagger} \right)\! + H_d.
\end{equation}
The part $H_d$ includes all the terms that are diagonal in the reference basis, such as the fermion mass term $m\sum_x (-1)^x \psi_x^{\dagger} \psi_x$, the gauge field energy $(g_n^2/2) \sum_x f(V_{x,x+1})$, and possibly proper counterterms. 

\begin{figure}
\centering
\includegraphics[width=0.85\textwidth]{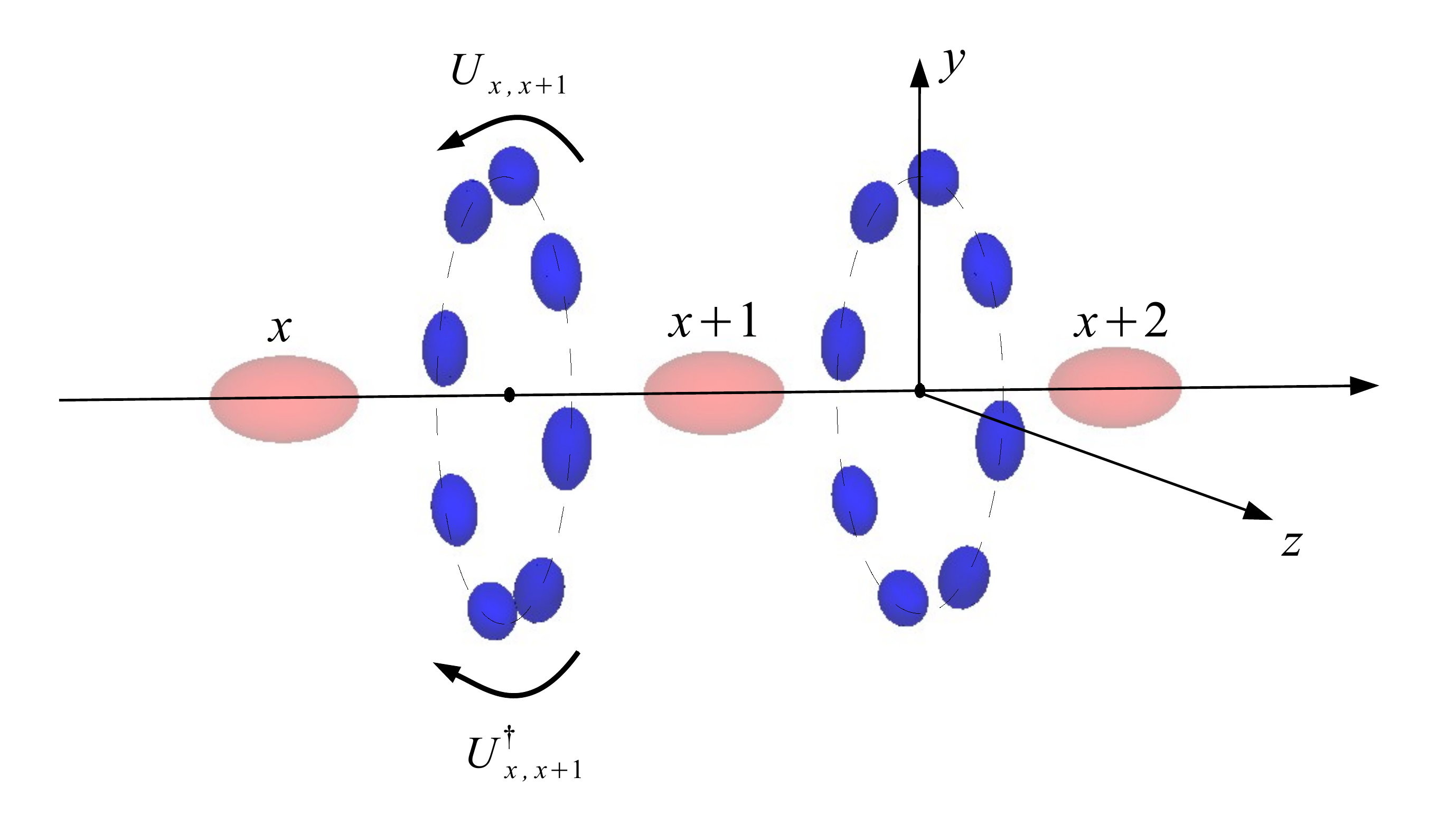}
\caption{Scheme of the physical implementation of the model. The fermions are confined in longitudinal lattice sites [light (red) spots], while the links are represented by ring-shaped transverse lattices, each one hosting one boson or fermion. The link particle is tightly confined in the longitudinal direction, while it can hop between nearest-neighbor sites [dark (blue) spots] on each ring lattice.}
\label{fig:lattice3d}
\end{figure}

The fermion hopping terms in (\ref{H0}) emerge naturally in condensed matter physics when a tight-binding approximation is assumed, in which only the lowest band of the lattice is energetically accessible. The matter field can be represented by a fermionic atomic species: the operator $\psi^{\dagger}_x$ creates an atom in the fundamental Wannier function centered on the lattice site $x$ \cite{bdz}. Tunneling between neighboring sites provides the mechanism for the hopping processes $\psi^{\dagger}_x\psi_{x+1}$. The staggered structure is obtained by modulating the depth of the lattice wells \cite{simul2}. A physical implementation of the action of the operators $U$ on links is less obvious, since it is first of all necessary to identify a proper system in which transitions between adjacent levels on a circle occur with the same amplitude (see Fig.~\ref{fig:1imm}). A possible implementation tool is represented by a longitudinal array of transverse ring-shaped lattices \cite{aoc}, with their axes aligned with $x$, each one representing a link and confining a single boson or fermion (the statistics being immaterial, since each ring contains one particle). The system is represented in Fig.~\ref{fig:lattice3d}. In the tight-binding regime, the fundamental Wannier states, centered on ring lattice sites, can be identified with the link reference basis states $|v_{x,x+1} \rangle$. Link particles are associated with the field operator $c_{(i) x,x+1}$, with $i=1,\dots,n$ the site index. Hopping between neighboring sites on the link yields the processes
\begin{equation}
U_{x,x+1} := \sum_{i=1}^n c_{(i+1) x,x+1}^{\dagger} c_{(i) x,x+1}, \quad U_{x,x+1}^{\dagger} := \sum_{i=1}^n c_{(i) x,x+1}^{\dagger} c_{(i+1) x,x+1},
\end{equation}
with $c_{(n+1)}\equiv c_{(n)}$. Transition between different rings are instead forbidden by a large energy barrier. Thus, the system and the hopping mechanism provide both the circular structure and the equality of transition amplitudes. Other possible implementations could involve a coupling between the internal levels of an atom confined on the link. However, in this case the unitary properties of transitions do not emerge by symmetry, and would require a fine tuning of the transition amplitudes \cite{cmr}.

While the term $H_d$ does not couple orthogonal eigenspaces of
$\Gamma_x$, each nondiagonal term in (\ref{H0}) maps a
state $|\phi\rangle$ of the physical space $\mathcal{H}_T= \ker{\Gamma}$
into another eigenstate
$|\phi'\rangle$ of each $\Gamma_x$ such that
\begin{equation}
\Gamma |\phi'\rangle = 2 \gamma_n |\phi'\rangle,
\end{equation}
where $\gamma_n=2(1-\cos(2\pi/n))$ is the first excited
eigenvalue of $\Gamma_x$, and the factor 2 is related to the fact
that each fermion hopping or Rabi transition on a link affects the
eigenvalue of $\Gamma_x$ in two neighboring sites. The
implementation strategy consists in adding to $H_{(0)}$ a term
which induces a large energy cost for going out of the physical
space, thus obtaining
\begin{equation}\label{H1}
H_{(1)} = H_{(0)} + \gamma_n^{-1} u\, \Gamma.
\end{equation}
If $u$ is much larger than the parameters appearing in $H_{(0)}$, the evolution of an initial state in the physical
subspace $\mathcal{H}_T$ is approximately given by the effective
Hamiltonian
\begin{equation}\label{Heff1}
H_{\mathrm{eff}} = P H_{(0)} P - P H_{(0)} Q {\left(
\gamma_n^{-1} u\, \Gamma\right)}^{-1} Q H_{(0)} P,
\end{equation}
with $P$ the projection operator on $\mathcal{H}_T$ and
$Q=\mathbb{I}-P$. Using the fact that $PH_{(0)}P=PH_d P$, and observing
that $H_{(0)}$ couples $\mathcal{H}_T$ only with the eigenspace of
$\Gamma$ belonging to 
$2\gamma_n$, Eq.~(\ref{Heff1}) simplifies to
\begin{equation}\label{Heff2}
H_{\mathrm{eff}} = P H_d P - (2 u)^{-1} P H_{(0)} Q H_{(0)} P.
\end{equation}
A straightforward computation of the second term yields, up to
immaterial constants, the final result
\begin{equation}\label{Heff3}
H_{\mathrm{eff}} = P \Bigl[ H_d - \frac{\tilde{t}
\tilde{w}}{u} \sum_x \!\left( \psi_x^{\dagger} U_{x,x+1}
\psi_{x+1} + \mathrm{H.c.} \right) +
\frac{\tilde{t}^2}{u} \sum_x n_x \!\left( 1 - n_{x+1} \right)
\Bigr] P,
\end{equation}
in which the correlated hopping terms in~(\ref{Ham})
appear. The counterterms in $H_d$ can be tuned to cancel the
undesired (though gauge-invariant) contribution in (\ref{Heff3}).
Despite the necessity to couple each site with the neighboring
links through $\Gamma_x$, this strategy can be more convenient
than a direct implementation of (\ref{Ham}), since the couplings
between sites and links are all diagonal in the reference basis. 
It should be
also noted that the operator $\Gamma$ can be replaced by any positive operator whose kernel is  
$\mathcal{H}_T$.

\section{Conclusions and outlook}

We have discussed a lattice
gauge model with a discrete $\mathbb{Z}_n$ Abelian symmetry. The model
represents an approximation to Wilson's lattice QED, that 
improves as the dimension $n$ of the link Hilbert space increases.
The obstacles to an experimental implementation of the model can
be overcome through the use of effective dynamics. Further
research will be devoted to two main avenues. The first one is the
identification of an atomic or condensed-matter system which is
suitable for an experimental implementation. 
The second is the application of the techniques developed in this Letter to non-Abelian gauge theories. Moreover, we shall scrutinize the convergence of the $\mathbb{Z}_n$ models towards $U(1)$, also in comparison with the QLM, and analyze in detail the phenomenology of the model, including chiral symmetry breaking and confinement.

\section*{Acknowledgements}
This work was partially supported by PRIN 2010LLKJBX on ``Collective quantum phenomena: from strongly correlated systems to quantum simulators,'' and by the Italian National Group of Mathematical Physics (GNFM-INdAM).

\appendix

\section{Weyl group representation in infinite dimensions}

Consider two Hermitian operators $A$ and $E$ with eigenvalue
equations
\begin{equation}\label{ae}
E|\epsilon\rangle = \epsilon |\epsilon\rangle , \qquad
A|\alpha\rangle = \alpha|\alpha\rangle ,
\end{equation}
and satisfying the canonical commutation relations
\begin{equation}\label{aecomm}
[E,A]= \ii \mathbb{I} .
\end{equation}
Observe that the above commutation relations make sense in an
infinite-dimensional space. Two operators satisfying Heisenberg's
commutation relations generate a representation of the Weyl group.
We are primarily interested in the unitary operators
\begin{equation}\label{mathcal}
\mathcal{U}(\eta) := \e^{-\ii\eta A}, \qquad
\mathcal{V}(\xi) := \e^{-\ii \xi E},
\end{equation}
with $\xi,\eta\in \mathbb{R}$, which act on the eigenstates (\ref{ae}) as spectral translations
(in opposite directions):
\begin{equation}\label{st}
\mathcal{U}(\eta) |\epsilon\rangle =
|\epsilon+\eta\rangle, \qquad \mathcal{V}(\xi)
|\alpha \rangle = |\alpha-\xi\rangle.
\end{equation}
Using the exponential form of the unitary operators
(\ref{mathcal}), the commutation relations (\ref{ae}) and the
Baker-Campbell-Haussdorf formula in the form $\e^{R}\e^{S} =
\e^{S}\e^{R}\e^{[R,S]}$, valid if $[R,S]$ is a multiple of the identity, one
obtains the property
\begin{equation}\label{VUVU}
\mathcal{U}(\eta) \mathcal{V}(\xi) = \e^{\ii \eta
\xi}   \mathcal{V}(\xi) \mathcal{U}(\eta) .
\end{equation}
The above result expresses
the noncommutativity of $E$ and $A$ at the level of the Weyl
group. The derivative of~(\ref{VUVU}) with respect to $\xi$ at $(\xi,\eta)=(0,1)$
yields
\begin{equation}\label{EU}
[E,U]=U,
\end{equation}
with $U:=\mathcal{U}(1)$, which is Eq.~(2) of the
Letter. Notice that by taking the derivative of~(\ref{VUVU}) with respect to $\eta$ and $\xi$ at the origin one also
reobtains Eq.~(\ref{aecomm}).

In the infinite-dimensional case (\ref{VUVU}) implies~(\ref{EU}). 
However, only (\ref{VUVU}) can be realized in 
the finite-dimensional  case (for discrete values of $\eta$ and $\xi$) by preserving the unitarity of all the
operators involved. By contrast, (\ref{EU}) does not admit a finite dimensional representation that preserves the properties of both $E$ and $U$.
Indeed, given two operators, $U$ unitary and $E$ Hermitian, on a
finite-dimensional space, the commutation relations~(\ref{EU})
cannot be satisfied, since they yield $UEU^{\dagger}=E-\mathbb{I}$,
which contrasts with $UEU^{\dagger}$ being isospectral to $E$ with a bounded spectrum.

In the Quantum Link model of QED, which is based on the
identification of the gauge degrees of freedom with spin variables
$E\to S_z$, $U\to S_{+} = S_x +i S_y$, the unitary operator $U$ is
replaced with a non-unitary one in order to mantain the commutation
relations $[E,U]=U$ valid. We will instead choose to maintain the
unitary structure of the gauge comparator $U$ in the
finite-dimensional case. To accomplish this task, we shall abandon
the Heisenberg's \textit{algebra} relations between the operators
$A$ and $E$, which cannot be realized in finite dimensions. We
will focus instead on the \textit{group} relations between their
complex exponentials (\ref{VUVU}), which admit a natural extension
to finite-dimensional spaces through the representation of the
Schwinger-Weyl group.

\section{The discrete Schwinger-Weyl group}

Let us consider an $n$-dimensional Hilbert space $\mathcal{H}_n$
and choose an orthonormal basis
$\left\{\ket{v_{\ell}}\right\}_{1\leq \ell\leq n}$. It is then
possible to define a unitary operator $U$ which rotates the basis
states as \cite{schwinger2001quantum}
\begin{equation}\label{W}
U \ket{v_{\ell}} = \ket{v_{\ell+1}} \quad \mbox{for } \ell<n, \qquad
U \ket{v_{n}} = \ket{v_{1}}.
\end{equation}
Since $U^n=\mathbb{I}$, the integer powers of $U$ constitute a
representation of the group $\mathbb{Z}_n$ of integers modulo $n$. Note
that the transition between the last and the first state of the
basis [last line of (\ref{W})] is necessary to ensure the
unitarity of $U$. The eigenvalue equation
\begin{equation}
U \ket{u_{k}} = \e^{-\frac{2\pi \ii}{n} k} \ket{u_{k}},
\end{equation}
with $k\in\{0,1,\dots,n\}$, is satisfied for
\begin{equation}\label{eigenW}
\ket{u_{k}} = \frac{1}{\sqrt{n}} \sum_{\ell=1}^n \e^{\frac{2\pi
\ii}{n} k \ell} \ket{v_{\ell}}.
\end{equation}
It is now possible to define an operator $V$ which rotates the
eigenbasis of $U$ as
\begin{equation}\label{V}
V \ket{u_{k}} = \ket{u_{k-1}} \quad \mbox{for } k>1, \qquad
V \ket{u_{1}} = \ket{u_{n}}.
\end{equation}
The operator $V$ is also unitary, with $V^n=\mathbb{I}$. From the
definition (\ref{V}) and the form (\ref{eigenW}) of the
eigenstates of $U$, one can easily demonstrate that
$\left\{\ket{v_{\ell}}\right\}_{1\leq \ell\leq n}$ is in fact
the eigenbasis of $V$, with
\begin{equation}
V \ket{v_{\ell}} = \e^{-\frac{2\pi \ii}{n} \ell} \ket{v_{\ell}}.
\end{equation}
The two operators $V$ and $U$ are thus called \textit{conjugated},
since each one rotates the other one's eigenbasis. It is also
relevant to observe that the action of $U$ and $V$ on each other's
eigenbasis {in~(\ref{W})-(\ref{V})} is the finite-dimensional counterpart of the spectral
translations (\ref{st}). (Notice the opposite signs.) 
The actions of $U$ and $V$ on a state do not commute. Indeed,
comparing
\begin{equation}
V U \ket{u_{k}} = \e^{-\frac{2\pi \ii}{n} k} V \ket{u_{k}} =
\e^{-\frac{2\pi \ii}{n} k} \ket{u_{k-1}}
\end{equation}
with
\begin{equation}
U V \ket{u_{k}} = U \ket{u_{k-1}} = \e^{-\frac{2\pi \ii}{n} (k-1)}
\ket{u_{k-1}},
\end{equation}
one obtains the relation
\begin{equation}\label{commut1}
U V= \e^{\frac{2\pi \ii}{n}}  V U,
\end{equation}
which can be immediately generalized to all the \textit{integer}
powers of the operators into \cite{schwinger2001quantum}
\begin{equation}\label{commut2}
U^{\ell} V^k  = \e^{\frac{2\pi \ii}{n} k \ell} V^k U^{\ell},
\end{equation}
coinciding with Eq.~(8) of the Letter. This result represents the
finite-dimensional generalization of (\ref{VUVU}). The most
striking difference is that the relation (\ref{commut2}) is valid
only for the \emph{discrete} set of integers, which implies that the
differentiation which leads from (\ref{VUVU}) to (\ref{EU}), is in
this case meaningless. However, the procedure leading to
(\ref{commut2}) is successful in preserving the unitarity of the
operators involved. Thus, (\ref{commut2}) is a valid starting
point for a gauge theory in which the local fields act on
finite-dimensional Hilbert spaces. The continuum limit is
recovered by introducing the Hermitian operators $A_n$ and $E_n$,
such that
\begin{equation}
U^{\ell} = \e^{-\ii \eta_{\ell} A_n}, \quad V^k=\e^{-\ii \xi_{k} E_n}
\end{equation}
with $\eta_\ell:= \ell\sqrt{2\pi/n}$ and $\xi_{k}:=k
\sqrt{2\pi/n}$. The Hermitian operators satisfy (\ref{aecomm}) 
by taking the limit $n\to\infty$, with $\eta_{\ell} \to \eta$ and $\xi_k\to\xi$~\cite{schwinger2001quantum}.

\end{document}